\documentstyle[12pt]{article}
\begin{document}
\def\headsep{0cm}
\def\headheight{0cm}
\def\topmargin{0cm}
\textheight 24cm
\newcommand{\lb}{\label}
\newcommand{\be}{\begin{equation}}
\newcommand{\ee}{\end{equation}}
\newcommand{\beqa}{\begin{eqnarray}}
\newcommand{\eeqa}{\end{eqnarray}}
\newcommand{\fr}{\frac}
\newcommand{\D}{\delta}
\newcommand{\ba}{\begin{array}}
\newcommand{\ea}{\end{array}}
\newcommand{\sig}{\sigma}
\newcommand{\del}{\partial}
\newcommand{\wv}{\wedge}
\newcommand{\al}{\alpha}
\newcommand{\la}{\lambda}
\newcommand{\La}{\Lambda}
\newcommand{\ep}{\epsilon}
\newcommand{\pr}{\prime}
\newcommand{\ti}{\tilde}
\newcommand{\om}{\omega}
\newcommand{\Omo}{\Omega}
\newcommand{\bi}{\bibitem}

\rightline{IC/93/94}

\rightline{May, 1993}

\vspace{2cm}

\begin{center}
{\large
A GENERAL SOLUTION OF THE BV-MASTER

\vspace{1cm}

EQUATION AND BRST FIELD THEORIES}

\vspace{2cm}

{\normalsize
\"{O}mer F. DAYI}

{\small \it
University of Istanbul,
Faculty of Science,  \\
Department of Physics,
Vezneciler, 34459 Istanbul, Turkey, \\
and \\
International Centre for Theoretical Physics,\\
 P.O. Box 586, 34100-Trieste, Italy}

\vspace{2.5cm}

\end{center}
{\small

For a class of first order gauge theories it was shown that
the proper solution of the BV-master equation can be obtained
straightforwardly.
Here we present the general condition which
the gauge generators
should satisfy to conclude
that this construction is relevant.
The general procedure is illustrated by its application
to the Chern-Simons theory
in any odd-dimension. Moreover, it is shown that
this formalism is also applicable to  BRST field theories,
when one replaces the role of the exterior derivative with
the BRST charge of  first quantization.
}

\pagebreak

\section{Introduction}

The Batalin -Vilkovisky (BV) method \cite{BV} which is a
systematic way of quantizing reducible gauge theories
is based on finding out the proper solution
of the (BV-) master equation. Unfortunately, obtaining
this solution is usually a tedious task. In ref. \cite{o1}
(which will be denoted as I) we
presented a procedure of discovering
the desired solution straightforwardly
for some  first order gauge theories, by utilizing
the general grading concept which was revealed  in the
study of topological quantum field theories \cite{cc}-\cite{ike2}
(for a review see ref. \cite{bbrt}).
In I it was only
shown  that the method is suitable for a very restricted
class of actions.

Here it is  demonstrated that,
for an action whose kinetic term is bilinear
in fields and linear in derivatives,
to know if the gauge generators
are related to the action in a certain way is sufficient
to conclude that the method of I is applicable.
Thus,  one avoids to check explicitly
(i.e. by using the original fields, ghosts, etc.), if the action
found in terms of the generalized fields satisfies
the master equation.  By inspection of the
action and the generators which are the main ingredients of the
gauge theories, one can easily determine if the
generalized field method of I is relevant
to find out the proper solution of the master equation.
In Section 3 as an example we study
the  application of the general method to the
generalization of the Chern-Simons theory to any
odd dimension introduced in
ref. \cite{mp}.

In Section 4 it is shown that, by replacing the role of the
exterior derivative with the first quantization BRST charge,
the general formalism is suitable to quantize BRST field theories.
This permits to obtain readily the proper solution of the
master equation for the string field theories.

\section{General Formalism}

The original and the ghost fields of a gauge theory
can be treated on the same footing, by
generalizing the exterior derivative, d,  as
\be
\lb{dtil}
\ti{d} \equiv d + \delta_B ,
\ee
where $\delta_B $ denotes the BRST transformation\cite{cc}.
Extension of the ordinary grading to include also
the ghost number is revealed to be useful in
writing the solution of the master equation
in a compact form for the topological quantum field
theories\cite{cc}-\cite{ike2}.

In order to utilize  the generalized grading to obtain  the
solution of the master equation we act according to the following
procedure.

If the original gauge theory is not already first order
in $d$, and the terms containing $d$ are not bilinear in fields,
one should find an equivalent formulation of it possessing these
properties. Thus we deal with the Lagrangian (action)
\be
\lb{ola}
L(A,B)= BdA + V(A,B),
\ee
which is supposed to be invariant under the gauge transformations
\be
\lb{3}
\D^{(0)} (A,B) = R^{(0)} (A,B)\la .
\ee

The minimal ghost content of the theory can be figured out  by
analysing the reducibility  of the  gauge transformations
(\ref{3}). Then, generalize the
original fields to include also the ghosts and antifields which
possess the same grading  with the  original ones
in terms of $\tilde{d}$, and substitute the original fields
$A,B$, with the generalized ones
$\ti{A},  \ti{B}  $, in the Lagrangian, (\ref{ola}).
The resulting action
\be
\lb{mass}
S \equiv L(\ti{A}, \ti{B})= \ti{B}d\ti{A} +V(\ti{A},\ti{B} ),
\ee
is invariant under the transformations generated by
\be
\lb{ggg}
\ti{R}(\ti{A},\ti{B})\equiv R^{(0)}(\ti{A} , \ti{B} ).
\ee

Of course, one should also generalize the gauge parameter,
$\la$, as $\ti{\la}$ whose generalized grading is the same
with the grading of $\la$.
To discover  the conditions which $\ti{R}$ should
fulfil to guarantee
that $S$ satisfies the  master equation, let us first examine
the invariances of $S$ which would yield that it is a solution
of the master equation.

We choose the signs of the field and antifield contents of
$\ti{A}$ and $\ti{B}$ as
\[
\ti{A}_A=(\phi_a ,\phi_i^*),\      \ti{B}_A=(-\phi_a^*, \phi_i),
\]
so that  under the transformations
\be
\lb{bt}
\D_B \ti{A}_A =\fr{\del_r S}{\del \ti{B}_A},\
\D_B \ti{B}_A =-\fr{\del_r S}{\del \ti{A}_A},
\ee
variation of $S$ will be
\[
\D_BS =(S,S)\equiv
2\fr{\del_r S}{\del \ti{B}_A}\fr{\del_l S}{\del \ti{A}_A}.
\]
If $S$ satisfies the master equation
\be
\lb{mas}
(S,S)=0,
\ee
(\ref{bt}) will
define the BRST transformations in accordance with the BV formalism.

By taking the derivatives of  (\ref{bt}),
one can define the transformations
\beqa
\D \ti{A}_A & = & -\fr{\del_l\del_rS}{\del \ti{B}_A \del \ti{A}_B}
\ti{\la}_1^B  - \fr{\del_l\del_rS}{\del \ti{B}_A \del \ti{B}_B}
\ti{\la}_2^B ,  \nonumber \\
\D \ti{B}_A & = & \fr{\del_l\del_rS}{\del \ti{A}_A \del \ti{A}_B}
\ti{\la}_1^B  +\fr{\del_l\del_rS}{\del \ti{A}_A \del \ti{B}_B}
\ti{\la}_2^B ,  \lb{ggi}
\eeqa
where $\ti{\la}_{1,2}$ are some parameters. Variation of
$S$ under (\ref{ggi}) can be shown to yield
\be
\D S = \fr{\del_r (S,S)}{\del \ti{A}_B} \ti{\la}_1^B
+\fr{\del_r (S,S)}{\del \ti{B}_B} \ti{\la}_2^B .
\ee
Thus if $S$ is invariant under the transformations (\ref{ggi})
with $\ti{\la}_1\neq 0$, $\ti{\la}_2 \neq 0$, one can conclude that
it satisfies the master equation, (\ref{mas}).

Fortunately, in  some circumstances to show that
$S$ satisfies the master equation, it is sufficient to know whether
$S$ is invariant under (\ref{ggi}) even if one of the
parameters $\ti{\la}_{1,2}$, is vanishing:
suppose that $S$ is invariant under the transformations
(\ref{ggi}) with the parameters
$\ti{\la}_1 \neq 0$, $\ti{\la}_2=0$. Thus, we may only
observe that $(S,S)$ is independent of $\ti{A}$.
$(S,S)$ possesses $(0,1)$ grading
(the first number indicates the usual
grading and the other one denotes the ghost number).
If the gradings of
the components of $\ti{B}$  are such that
it is impossible to construct a function possessing $(0,1)$ grading
only in terms of $\ti{B}$, we can conclude that
$(S,S)$ vanishes. The other case, $\ti{\la}_1 = 0$, $\ti{\la}_2\neq 0$,
can be examined similarly.

Therefore, $S$ which is obtained by substituting the original
fields with the generalized ones in the original action so that
invariant under the transformations generated by (\ref{ggg}),
is a  solution of the master equation,
(\ref{mas}), if $\ti{R}$ generates
the transformations (\ref{ggi}) where both of the parameters
are non-vanishing or one of them vanishes but
$(S,S)$ cannot depend on the field related to the vanishing
parameter due to its grading.

By construction $S(\ti{A},\ti{B} )$ possesses the correct classical
limit. Moreover, $\ti{A}$ and $\ti{B}$ include
all the fields of the minimal sector and
because of the form of $S$, (\ref{mass}),
\[
rank\  \left| \fr{\del^2 S}{\del (\ti{A},\ti{B}) \del
(\ti{A},\ti{B})} \right| = N,
\]
where $N$ is the number of the components of
$\ti{A} $ or $\ti{B} $. Hence, we conclude that
under the above mentioned conditions $S(\ti{A},\ti{B})$
is the proper solution of the master equation.

\section{Chern-Simons theory in $d=2n+1$}

The theories studied before in refs. \cite{o1}-\cite{ike2},
can be shown to satisfy the above conditions.
Nevertheless to illustrate the method
we would like to
study  the Chern-Simons theory in any $d=2n+1$, which
is given with the action (we suppress Tr)
\be
\lb{cs}
L_d= \fr{1}{2} \int_{M_d} \left( A \wedge dA
+\fr{2}{3} A\wedge A\wedge A \right) .
\ee

When we define
\[
A=\phi +\psi \equiv
\sum_{i=0}^{n-1} \phi_{2i+1}+\sum_{i=0}^n \psi_{2i},
\]
where   $\phi$ and $ \psi $ are  Lie-algebra valued, respectively,
bosonic $2i+1$-form and fermionic $2i$-form. it reads
\be
\lb{sl}
L_d  =   \fr{1}{2} \int  _{M_d} \left( \phi \wedge d \phi
 +\fr{2}{3} \phi \wedge \phi \wedge \phi +
\psi \wedge D_\phi \psi \right),
\ee
where $D_\phi \equiv d +[ \phi ,\  ]$.
(\ref{sl}) is followed from the fact
that in the integral only the terms possessing odd grading
survive.
In (\ref{sl}) one recognizes the theory introduced in
ref. \cite{mp}\footnote{Although quantization of this
theory in terms of the BV method is already considered in
ref. \cite{bro2} by making use of the generalized grading concept,
our discussion differs from that drastically. Neither the ghost
fields given in (\ref{ft}) nor the final answer (\ref{fa})
are in accord
with ref. \cite{bro2}. In ref. \cite{bro2} three different ghost numbers
are introduced, thus  in the final answer  (\ref{fa}),
it is claimed that
$\int \ti{\psi} \ti{\phi}\ti{\psi} = \int \ti{\psi} \phi \ti{\psi}$,
which does not yield the correct BRST transformation
of $\ti{\psi}$,  which should include also the contributions
coming from $\La$, (\ref{116}).}
(see \cite{dmg} for a supersymmetric formalism of the BF theory
which possesses some common features).

The action (\ref{sl}) is invariant under
\be
\lb{rgt}
\D A= d\Sigma + [ A,\Sigma ] ,
\ee
where
\be
\lb{116}
\Sigma = \La +\Xi \equiv \sum_{i=0}^{n-1} \La_{2i}
+\sum_{i=0}^{n-1} \Xi_{2i+1}.
\ee
$\La$ and $\Xi$ are bosonic and fermionic, respectively.
(\ref{rgt}) are reducible on mass shell. Indeed, when the equations
of motion
\[
F_\phi - \psi^2=0,\     D_\phi \psi =0,
\]
are satisfied, one can see that
\[
Z_m Z_{m+1}=0,\ m=0, \cdots 2n-2,
\]
where $Z_0$ is the gauge generator and
\[
Z_{2m}=\left(
\ba{cc}
D_\phi &  \psi \\
\psi   &  D_\phi
\ea
\right) ,
Z_{2m+1}=\left(
\ba{cc}
D_\phi &  -\psi \\
-\psi   &  D_\phi
\ea
\right) .
\]

By examining the reducibility properties one introduces ghosts
and ghosts of  ghosts fields, the related antifields and obtains
\beqa
\ti{\phi}  &  =  &
\sum_{i=0}^{n-1}  \left[ \phi_{(2i+1,0)}
+ \sum_{j=1}^{2i+1} \eta_{(2i+1-j,j)}  +\phi^*_{(2i+2,-1)}
+ \sum_{j=-2n+2i+4}^{-2} \eta^*_{(2i+1-j,j)} \right]  \nonumber \\
 & &  \lb{ft} \\
\ti{\psi} &  =  &
\sum_{i=0}^n  \psi_{(2i,0)}
+\sum_{i=1}^n \sum_{j=1}^{2i} \kappa_{(2i-j,j)}
+\sum_{i=0}^n \psi^*_{(2i+1,-1)}
+ \sum_{i=0}^{n-1} \sum_{j=-2n+2i+1}^{-2} \kappa^*_{(2i-j,j)}  .\nonumber
\eeqa
The antifield of the field $a_{(k,l)}$ is defined as $a^*_{(2n+1-k,-l-1)}$.
Observe that  $\ti{\phi}$ and $\ti{\psi}$ are,
respectively,  collection of $2i+1$-forms
and  $2i$-forms. Now, in terms of
$\ti{A} =\ti{\phi}+\ti{\psi}$, we can write
\be
\lb{sd}
S_d=\fr{1}{2}\int  _{M_d} \left( \ti{A} d \ti{A}
+ \fr{2}{3} \ti{A}^3 \right) .
\ee
$S_d$ is a proper solution of the master equation,
because it is invariant under the transformations generated by
\[
\ti{R}=d+[ \ti{A}, ]=\fr{\del^2 S_d}{\del \ti{A}^2},
\]
following as the generalization of (\ref{rgt}).
Because of the sign assignments in (\ref{ft}) the transformations
(\ref{ggi}) are given as
\[
\D \ti{A}_A=\omega_{AB}\fr{\del_l \del_rS}{\del \ti{A}_B \del
\ti{A}_C}\ti{\Sigma }_C;\ \omega_{AB}=\left(
\ba{cc}
{\bf 0} & {\bf 1} \\
- {\bf 1} & {\bf 0}
\ea
\right) ,
\]
where the generalized gauge parameter
\beqa
\ti{\Sigma} & = & \ti{\La} + \ti{\Xi}, \nonumber  \\
\ti{\La} &  =  &
\sum_{i=0}^{n-1}  \La_{(2i,0)}
+\sum_{i=1}^{n-1} \sum_{j=1}^{2i} \la_{(2i-j,j)}
+\sum_{i=0}^{n-1} \La^*_{(2i+1,-1)}
+ \sum_{i=0}^{n-2} \sum_{j=-2n+2i+1}^{-2} \la^*_{(2i-j,j)}  \nonumber \\
\ti{\Xi}  &  =  &
\sum_{i=0}^{n-1}  \left[ \Xi_{(2i+1,0)}
+ \sum_{j=1}^{2i+1} \xi_{(2i+1-j,j)}  +\Xi^*_{(2i+2,-1)}
+ \sum_{j=-2n+2i+4}^{-2} \xi^*_{(2i+1-j,j)} \right] . \nonumber
\eeqa

This example is somehow different from the general case, because
the general gradings of the components of $A$ are not the same.
But the integral selects only the terms with the correct grading.
One could write the solution of the master equation by using the
generalized forms each of which possessing only one grading,
and then gather them to obtain (\ref{sd}).
In terms of $\ti{\phi}$ and $\ti{\psi}$ components
(\ref{sd}) is given as
\be
\lb{fa}
S_d = \int  _{M_d} \left( \ti{\phi} d \ti{\phi} +\fr{1}{3}\ti{\phi}^3
+\ti{\psi}(d+\ti{\phi} )\ti{\psi} \right) ,
\ee
by observing: $i) $
$\int  _{M_d}\ti{\phi}d\ti{\psi}=0$,
integral is defined to possess zero ghost number so that one of
the two factors of the integrand
should be a field and the other one an antifield,
which belong to the same reducibility level, but then
it is an even form,
$ii)$ $\int  _{M_d} \ti{\phi}^2\ti{\psi}=0$,
integrand will be
a $M$-form, where $M=2i+1-j+2m+1-k+2p+1-q$, but also due to
condition on the ghost number we have $j+k+q=0$, therefore $M=$even,
iii) $\int  _{M_d} \ti{\psi}^3=0$, integrand will be
a $K$-form, where $K=2i-j+2m-k+2p-q$, and again  due to
fact that its ghost number should vanish,
we have $j+k+q=0$, therefore $K=$even.
Hence their integrals on $d=2n+1$-manifold vanish.

\section{BRST Field Theory}

When we deal with a constrained hamiltonian system (gauge theory)
its first quantization will yield  the BRST charge $Q$,
which is fermionic and nilpotent, $Q^2=0$. To define
this charge one should introduce first quantization ghosts,
and associate them the
``algebraic ghost
number", $N_a$. By construction  $Q$ possesses $N_a(Q)=1$.
The BRST charge, $Q$, acts on the fields which are valued in
the Hilbert space of the first quantized theory, and the physical
states are defined  as
\[
Q \chi_p =0,\ \chi_p \neq Q(...).
\]
The fields can be written as functionals or in terms of the normal
modes. In general we can attribute algebraic
ghost numbers to the fields and
define an inner product to write a free action whose equations of
motion are the physical state conditions\cite{sie}.
Inner product can be allowed to
carry an algebraic ghost number or not. We deal with an inner product
which does not carry algebraic ghost number. Then the
free action is
\be
\lb{la}
L_0(\chi , \bar{\chi})=\bar{\chi} Q\chi ,
\ee
where ($\ep$ denotes the Grassmannn parity)
\beqa
N_a(\bar{\chi})=-1, &  N_a(\chi)=0, \nonumber \\
\ep (\bar{\chi})=1, & \ep (\chi )=0.         \nonumber
\eeqa
Because of the nilpotency
of $Q,$ (\ref{la}) is invariant under the gauge transformations
\be
\lb{qgi}
\D \chi =Q \La_1,\  \D \bar{\chi} =  \bar{\La}_1Q ,
\ee
where the gauge parameters possess
\[
N_a(\La_1)=-1 ,\  N_a(  \bar{\La}_1)=-2 .
\]
This algebraic ghost number assignment follows from
the fact that we would like to interpret the first quantization
BRST charge, $Q$, as the exterior derivative, though
in this case there are ``negative forms" ($\equiv$
negative $N_a$ states).
For the sake of generality, we suppose that any integer
algebraic ghost number is available. Obviously, the gauge invariance
(\ref{qgi}) is infinitely reducible:
\beqa
\D \La_n =Q\La_{n+1}, & \D  \bar{\La}_n=  \bar{\La}_{n+1}Q  \nonumber \\
N_a(\La_n)=-n, &  N_a(  \bar{\La}_n)=-(n+1),  \lb{agn} \\
\ep (\La_n)=n,  &  \ep (  \bar{\La}_n)=n+1.   \nonumber
\eeqa
To perform the BV quantization we need to introduce the ghost and
the ghost of ghost fields, $\eta_n,\  \bar{\eta}_n $,
which possess the following ``gauge
ghost number", $N_g$ (of course, gauge ghost number of
$\chi$ and $ \bar{\chi} $ is zero),
\[
N_g(\eta_n)=N_g(  \bar{\eta}_n )=n,\   \ep (\eta_n)=\ep (\La_n )+1,\
\ep (  \bar{\eta}_n)=\ep  (  \bar{\La}_n)+1.
\]
They also carry algebraic ghost number according to (\ref{agn}).

Now, we can use the formalism given in Section 1, by the replacement
\[
d  \rightarrow  Q,\  \D_B \rightarrow \Delta_B ,
\]
where $\Delta_B$ denotes the BRST transformation resulting from the
second quantization.
We can attribute a generalized grading to the fields introduced
above: $\chi,\ \eta_n$ possess $0$-generalized grading and
$ \bar{\chi} ,  \bar{\eta}_n$ possess $-1$-generalized grading. In  the BV
quantization scheme we should also introduce the related antifields.
Antifields are defined such that the sum of the total ghost numbers
of a field and its antifield should be $-1$. Thus they are given as
\[
\chi^*_{(0,-1)},\ \bar{\chi}^*_{(1,-1)},\ \eta^*_{n\ (n, -n-1)},\
 \bar{\eta}^*_{n\ (n+1, -n-1)},
\]
where the first number in the parenthesis indicates the
algebraic ghost number and the second denotes the gauge
ghost number.

According to the rules given in Section 1 we can
group the fields which possess the same generalized grading:
\beqa
 \ti{\chi} & = & \chi_{(0,0)} +\eta_{n\ (-n,n)} +
 \bar{\chi}^*_{(1,-1)} +  \bar{\eta}^*_{n\ (n+1, -n-1)}, \nonumber \\
 \ti{\bar{\chi}} &  =  &
-\chi^*_{(0,-1)}- \eta^*_{n\ (n, -n-1)}+
\bar{\chi}_{(-1,0)} +  \bar{\eta}_{n\ (-n-1,n)}. \nonumber
\eeqa
Then we generalize the original action, (\ref{la}), as
\be
\lb{lam}
S_0=L_0(\ti{\chi},\ti{\bar{\chi}}) = \ti{\bar{\chi}} Q  \ti{\chi} ,
\ee
where the multiplication is defined such that $\ep (S_0)=0$,
and $N_a(S_0)=N_A(S_0)=0$. One can easily check that $S_0$ satisfies
the master equation by observing that it is invariant under
\[
\D  \ti{\chi} =\fr{\del_l S_0}{\del \ti{\bar{\chi}} } ,\
\D \ti{\bar{\chi}} =\fr{\del_r S_0}{\del  \ti{\chi} } ,
\]
and it is also proper:
\[
rank \left[ \fr{\del^2 S_0}{\del  \ti{\chi} \del \ti{\bar{\chi}}}\right] =
\#\ of\  \ti{\chi} =\#\ of\ \ti{\bar{\chi}} .
\]

To expose  the resemblance with the general formalism introduced in
Section 1 and for future use,
observe that the original gauge
invariance (\ref{qgi}), can be written as
\be
\lb{oqi}
\left(
\ba{c}
\D \chi \\
\D \bar{\chi}
\ea
\right)
=R^{(0)}(L_0,\chi ,  \bar{\chi} )
\left(
\ba{c}
\La_1 \\
 \bar{\La}_1
\ea
\right)
\equiv
\left(
{\large
\ba{cc}
\fr{\del_l \del_r L_0}{\del   \bar{\chi} \del \chi } &
\fr{\del_l \del_r L_0}{\del   \bar{\chi} \del   \bar{\chi}}  \\
 \fr{\del_l \del_r L_0}{\del \chi \del \chi}  &
-\fr{\del_l \del_r L_0}{\del \chi \del   \bar{\chi}}
\ea  }
\right)
\left(
\ba{c}
\La_1 \\
\bar{\La}_1
\ea
\right).
\ee
Thus the generalized action $S_0$ will be invariant under
the transformations generated by
\be
\lb{rg}
\ti{R}(S_0) \equiv R^{(0)}( S_0, \ti{\chi} , \ti{\bar{\chi}} ).
\ee
This is equivalent to the transformation (\ref{ggi}),
used in Section 1
to show that $S_0$ satisfies the master equation.

Interactions can be introduced in terms of some vertex operators
$V^{(k)}_N$, as
\be
\lb{x}
L_I(\chi ,\bar{\chi})= \sum_N \sum_{k=1}
 \bar{\chi} (1) \cdots  \bar{\chi} (N-k) V_N^{(k)}
\chi (N-k+1) \cdots \chi (N),
\ee
where $L_I$ is defined to be bosonic and to possess zero ghost numbers.
The ranges of $N$ and $k$ depend on the requirements.

There are mainly
two ways of specifying the interaction terms: $i)$ keep the gauge
invariance as it was for the non-interacting case (\ref{qgi}),
by demanding $QV_N=0$\cite{hue}, or ii) require that the gauge invariance
is generated by the generators
$R^{(0)}(L,\chi ,\bar{\chi})$, where $R^{(0)}$ is defined
in (\ref{oqi}), and
\[
L(\chi,\bar{\chi})=L_0+L_I.
\]
We deal with the latter case which allows us to obtain
the proper solution of the master equation
as before:
\[
S=L(\ti{\chi},\ti{\bar{\chi}});\     (S,S)=0.
\]

BRST field theory is suitable to construct string field theories, and
most of the string theories studied earlier can be seen to obey
the above conditions (e.g. see refs. \cite{nc}-\cite{tho}).
Recently this is observed in ref. \cite{zwi} for the closed
bosonic string field theory. There the inner product
carries  algebraic ghost
number, so that the original action depends
only on one field
$\chi$. Nevertheless,
one can see that all of the conditions of the general
procedure introduced above are
satisfied, hence one can directly obtain the
proper solution of the master
equation by replacing the original field with the one
including all of the ghost fields
and the antifields (compare with the effort spent in ref.
\cite{zwi} to show that indeed this generalization yields the proper
solution of the master equation).

To perform gauge fixing one introduces some new fields and
Lagrange multipliers, which can be grouped as
$\ti{C}$ and $\ti{\Pi}$ with $0$- and $1$-generalized grading, respectively.
Moreover in terms of the gauge fixing fermion $\ti{\Psi}$
which possesses $-1$-generalized grading one fixes
the gauge freedom as $\phi^* =\del \ti{\Psi}/\del \phi$. In the enlarged
space the proper solution of the master equation is
\[
S_e=S+\ti{C}^* \ti{\Pi}.
\]
Let us deal with $k=1$ in (\ref{x}) and
choose the gauge fixing fermion as
\[
\ti{\Psi}=\ti{C}(\ti{\bar{\chi}} -\ti{\chi} O ),
\]
where $O$ is any operator
possessing $N_a(O)=-1$, depending on the  first quantization variables.
First of all, this gauge fixing fermion renames the antifields
which are present in $\ti{\chi}$ as antighosts.
In the related path integral
after integrating  over the Lagrange multipliers, $\ti{\Pi}$ ,
the gauge fixed action will be
\be
\lb{gfb}
S_g=\ti{\chi} OQ  \ti{\chi} +
 \sum_N  O,V_N \ti{\chi} (1) \cdots  \ti{\chi} (N) .
\ee

This procedure suggests that one can proceed in the
reverse direction: suppose that
in the superspace given in terms
of the first quantization variables
there is an action in the form (\ref{gfb}).
It may be possible to construct a gauge invariant action
wich leads to it after gauge fixing, by finding
the appropriate $Q$ and $O$ operators.

To elucidate the above procedure we briefly discuss its application
to the relativistic particle
(see also ref.\cite{hol}),
which is defined in terms of the
BRST charge
\[
Q=c(p_\mu^2 +m^2),
\]
where $c^2=0$ and $p_\mu =-i\del / \del x^\mu$.
In this case the available algebraic ghost
numbers are $(-1,0,1)$, so that the linear gauge invariance is
\[
\D \chi =Q\La,\  \D \bar{\chi}=0,
\]
and moreover, it is irreducible.
Hence the generalized fields are
\beqa
 \ti{\chi} & = & \chi_{(0,0)} +\eta_{ (-1,1)} +
 \bar{\chi}^*_{(1,-1)}
, \nonumber \\
 \ti{\bar{\chi}} &  =  &
-\chi^*_{(0,-1)}- \eta^*_{(1, -2)}+
\bar{\chi}_{(-1,0)} . \nonumber
\eeqa

In terms of the vertex operator
\[
V_N(1, \cdots , N)=c
\D (x_1-x_2)\cdots \D (x_{N-1}-x_N),
\]
one can write the action
\be
\lb{act}
L_r(\chi ,\bar{\chi} )
= -\int d^dx \bar{\chi}Q\chi + \int \prod_{n=1}^N d^dx_n \sum_{N=3}
 \bar{\chi} (1)  V_N
\chi (2) \cdots \chi (N),
\ee
which can easily be seen  to be invariant under the transformations
generated by $R^{(0)}(L_r, \chi ,\bar{\chi})$
(see (\ref{oqi})). Because of the algebraic
ghost number restrictions
$\del^2 S_I / \del \bar{\chi}^2$ should vanish,
so that the action given in (\ref{act}) is the general one.

When one generalizes the action $L_r$ as
\[
S_r=L_r(\ti{\chi}, \ti{\bar{\chi}}),
\]
its invariance under the transformations generated by
$\ti{R}(S_r)$ yield that $(S_r,S_r)$ is independent of
$\ti{\chi}$ only, because $\bar{\La}_1=0$. Obviously
$(S_r,S_r)$  cannot depend only on $\ti{\bar{\chi}}$, so that
it vanishes.

Let us choose the following gauge fixing fermion
after enlarging the space of fields as described above,
\[
\ti{\Psi} = \ti{\kappa} (\ti{\bar{\chi}} -\ti{\chi} p_c),
\]
where $\{ p_c ,c\} =1$. The effective gauge fixed
action will be
\[
S_{rg}= \int d^dx \left( \ti{\chi}[ \del^2 -m^2]\ti{\chi} + \sum_{N=3}
 \ti{\chi}^N \right),
\]
where $\bar{\chi}^*$ should be renamed as antighost.

\vspace{.1in}

\begin{center}
{\bf \large Acknowledgments}
\end{center}

I am grateful to G. Thompson for helpful discussions.

I thank
Professor Abdus Salam, the International
Atomic Energy Agency and UNESCO for hospitality at the ICTP.

This work is partially supported by the Turkish Scientific and
Technological Research Council (TBTAK).

\pagebreak

\end{document}